\documentclass[5p,times]{elsarticle}

\usepackage{ecrc}

\volume{00}

\firstpage{1}

\journalname{Astronomy and Computing}

\runauth{M. Stănescu and O. Văduvescu}

\jid{A&C}

\jnltitlelogo{Astronomy and Computing}

\usepackage[utf8]{inputenc}
\usepackage{tabularx}
\usepackage{siunitx}
\usepackage{multicol}
\usepackage{geometry}
\usepackage{url}
\usepackage{subcaption}
\usepackage{natbib}

\renewcommand{\elslogo}{}
\renewcommand{\sdlogo}{}

\begin{document}
	
	\begin{frontmatter}
		
		

		\dochead{}
		
		\title{The Umbrella software suite for automated asteroid detection}
		
		\author[AB,KCL]{M.~Stănescu}\ead{malin.stanescu@gmail.com}
		\author[ING,IAC,UCV,AB]{O.~Văduvescu}\ead{ovidiu.vaduvescu@gmail.com}
		\address[AB]{Astroclubul București, Str. Cuțitul de Argint nr. 5, Sector 4, Bucharest, Romania}
		\address[KCL]{Department of Mathematics, King's College London, Strand, WC2R 2LS, London, United Kingdom}
		\address[ING]{Isaac Newton Group (ING), Apt. de correos 321, E-38700, Santa Cruz de La Palma, Canary Islands, Spain}
		\address[IAC]{Instituto de Astrofisica de Canarias (IAC), Via Lactea, 38205 La Laguna, Tenerife, Spain}
		\address[UCV]{University of Craiova, Str. A. I. Cuza nr. 13, 200585, Craiova, Romania}

		\begin{abstract}
			We present the Umbrella software suite for asteroid detection, validation, identification and reporting. The current core of Umbrella is an open-source modular library, called Umbrella2, that includes algorithms and interfaces for all steps of the processing pipeline, including a novel detection algorithm for faint trails. Building on the library, we have also implemented a detection pipeline accessible both as a desktop program (ViaNearby) and via a web server (Webrella), which we have successfully used in near real-time data reduction of a few asteroid surveys on the Wide Field Camera of the Isaac Newton Telescope. In this paper we describe the library, focusing on the interfaces and algorithms available, and we present the results obtained with the desktop version on a set of well-curated fields used by the EURONEAR project as an asteroid detection benchmark.
		\end{abstract}
		
		
	\end{frontmatter}
	
	
	
	\newpage

\section{Introduction}
\subsection{Previous work}
Pipelines for automating the detection and processing of moving objects have been proposed and implemented \citep{mops-Allekotte} in the past. Some, such as \citet{mops-Bekte}, have focused on detection only, with little subsequent processing. Due to the computationally intensive tasks associated with detection, more recently such systems have been implemented in the cloud \citep{mops-a24n}. Of these pipelines, very few (and none from the major surveys of minor planets, except Pan-STARRS \cite{Magnier_2020} \cite{panstarss_ipp}) are known to be open-source.\\

Among asteroid detection methods, synthetic tracking \citep{syntrack-2005} has the best detection properties, however, it requires significant processing power and fast camera readout, thus only recently \citep{syntrack-Shao_2014} has become accessible to surveys \citep{syntrack-zhai2018technical}. Some faster alternatives to synthetic tracking have been suggested by \citet{trails-brown}, albeit with worse noise rejection (and still requiring fast imaging equipment). Therefore, efforts were mainly concentrated on the blink method.\\

Among detection algorithms for the blink method, notable are maximum likelihood techniques, as described in \citet{dawson2016blind}, which, given a set of assumptions, yield an optimal algorithm. Such detection methods are non-trivial to implement, especially since image defects can be hard to model accurately, and the optimization nature of the task can make it computationally expensive compared to simpler algorithms, such as flood fill.\\

One project, \citet{Waszczak_2017}, targeting asteroid trails using the Palomar Transient Factory camera has focused on applying machine learning to a set of morphological features of the detections to separate the latter from noise. Notably, this project has documented well (through a rather extensive graphical enumeration) the failure modes of a simple detection pipeline. From the authors' experience, these occur on other detection pipelines too, but most can be solved through detection post-processing (the approach taken by Umbrella), significantly improving rejection rate of false positives and loss rate of true positives. However, this project uses the flood fill algorithm, which is less efficient at detecting trails, and the software is not known to be open-source.
\subsection{History of the project}
The version presented in this paper is an early preview of the third version of the Umbrella software. The Umbrella project started in June 2015, as a volunteer summer research project that has extended and expanded intermittently over the past 5 years, targeting automated asteroid detection using the blink method (requiring a minimum of 3 images per field). Umbrella is now part of the larger EURONEAR project\footnote{\url{www.euronear.org}}. The following is an approximate timeline of the development of the software.\\

At the start of the project, Source Extractor was considered for detecting light sources in the image; the catalogs generated were to be compared and only the transient objects were to be kept. The method proved to be difficult due to the large number of bad pixels and cosmic rays that plagued the images. It was decided that the moving object detection and processing software would use its own detection and filtering algorithms. Within the next year, the resulting desktop application, named SourceUmbrella, peaked at 80 \% detection rate (compared to human blink methods) on the sample images provided (from the EURONEAR archive \citep{euronear-discoveries1}\citep{euronear-discoveries2}) for development (training / development set) and reached 20-50 \% detection rate on new images (test set). However, it had much worse performance in detecting trailed objects (such as Near Earth Asteroids). It was also very rigid in accepting FITS files, which made the use of other image sources difficult. The architecture of the application also proved unsustainable for the wide array of variations in the pipeline.\\

Beginning in December 2017, the development resumed on a new design, an open-source modular library, called Umbrella2, that would be much more flexible and overcome the shortcomings of SourceUmbrella. In particular, due to the low detection rate of Umbrella2 on trailed objects, and taking into account that another similar software developed for the EURONEAR project, NEARBY \citep{nearby-conference}\citep{nearby-unpub}, would not detect trailed objects either, the development focused more on trail object detection. In that sense, in February 2018, a new line segment detection algorithm was designed, based on the Hough Transform \citep{hough-dudahart}, to detect long and faint NEA trails (SV-AFAV-HT, i.e. State Variable Auxiliary Feature Augmented Voting Hough Transform, paper by Stănescu in prep - see Annex A for a working description). By November 2018 a desktop application building on the Umbrella library was ready for processing images and had an early live test during the November 2018 EURONEAR survey; using the images with astrometry (more precisely, the FITS World Coordinate System headers) resolved by the NEARBY pipeline, as described in  \citet{pipeline-denisa} (using AstrOmatic software: Source Extractor \citep{sExtractor}, SCAMP \citep{scamp} and SWarp \citep{swarp-terapix}). Since then, many components of the library have been improved and many bugs have been resolved.\\

In the second half of 2019, the focus shifted to the development of a web-based deployment of Umbrella. This resulted in the Webrella\footnote{\url{http://s141.central.ucv.ro:49239/}} server, a web interface to the configurable pipeline also used by the desktop version. The web server was ready by the end of October (2019), and had been tested with images from previous runs. It also had a very limited live test during the October-November survey, using images captured with the Wide Field Camera (WFC) on the Isaac Newton Telescope (INT). Unfortunately, some crashes in the NEARBY pipeline (installed on the same server) that provided WCS-reduced images to Webrella hindered live testing of the web version during the next INT survey. Nevertheless, the desktop version (which acted as a backup) has been used successfully for near real-time in the survey, using the images from the remaining NEARBY server.
\subsection{Project source code and technology stack}
The entire Umbrella suite is developed on top of Mono / .NET Framework, with the library and the ViaNearby pipeline written entirely in C\#. The server backend for Webrella is built on top of the Nancy Framework, with plain HTML on the browser side. The desktop GUI is implemented using WinForms. The software should be portable across operating systems (Windows, Linux, MacOS, BSD) and architectures (x86, ARM, etc.), with minor caveats mostly depending on the underlying Mono/.NET implementation (for example WinForms is not implemented on 64-bit MacOS, so the desktop application is unavailable on that platform). The authors have tested it only on Windows, Linux and MacOS on x86. It should be noted that the custom quickselect algorithm (presented below in this paper) is by default compiled to require 64-bit support.\\

\begin{figure*}
	\centering
	\includegraphics[width=.9\linewidth]{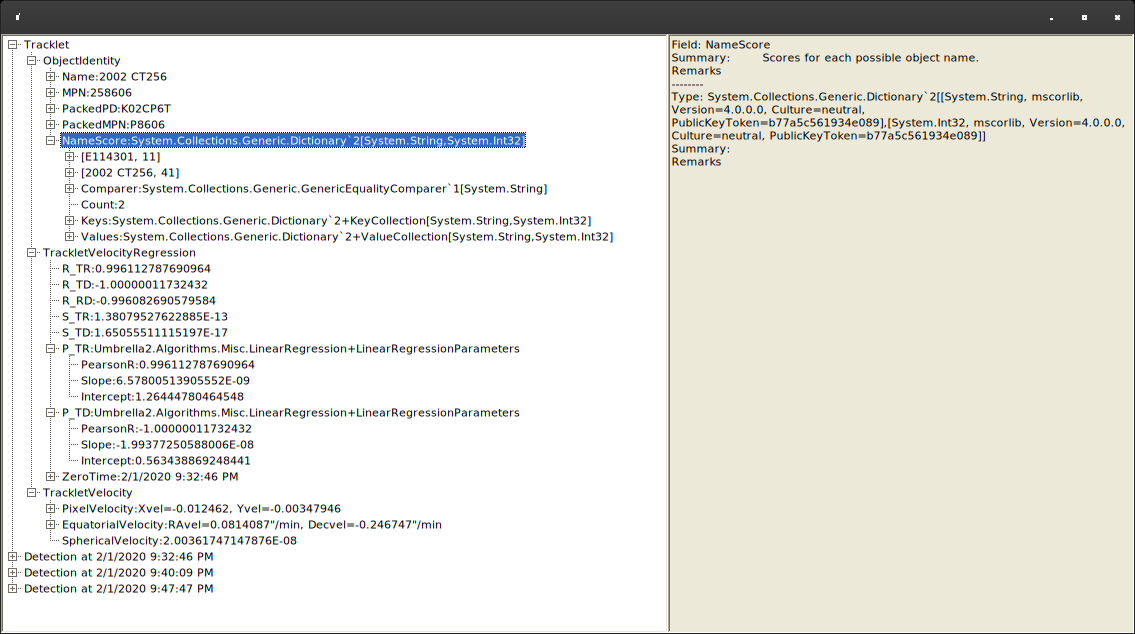}
	\caption[Tracklet properties]{Almost all information built by the pipeline regarding a given object is available in the property viewer. Explanations about the various properties and their fields are pulled directly from source code documentation (including user-supplied properties). Here, one may see how the SkyBoT name (2002 CT256, score 41) takes priority over the default provisional tracklet name (E114301, i.e. night 1, field 14, ccd 3, tracklet 1).}
	\label{fig:vn-objprop}
\end{figure*}

The source code of the core library is available on Github, at \url{https://github.com/mostanes/umbrella2}. The code is documented in-line using XML, covering almost all classes, properties, fields and methods, both public and private. This documentation is available both to IDEs (typically available when writing the code, through a similar mechanism to autocomplete) and Umbrella itself (for example, when inspecting internal data structures with the property inspector, see Figure \ref{fig:vn-objprop}). Stable versions are available on NuGet\footnote{\url{https://www.nuget.org/packages/umbrella2/}}. The source code of ViaNearby is available at \url{https://github.com/mostanes/umbrella2-euronear}. The ViaNearby pipeline code has scarce in-line documentation. Besides the documentation available in the source code, releases of the ViaNearby pipeline come with usage and reference manuals.
\begin{figure*}
	\begin{subfigure}{\textwidth}
	\centering
	\includegraphics[width=.99\linewidth]{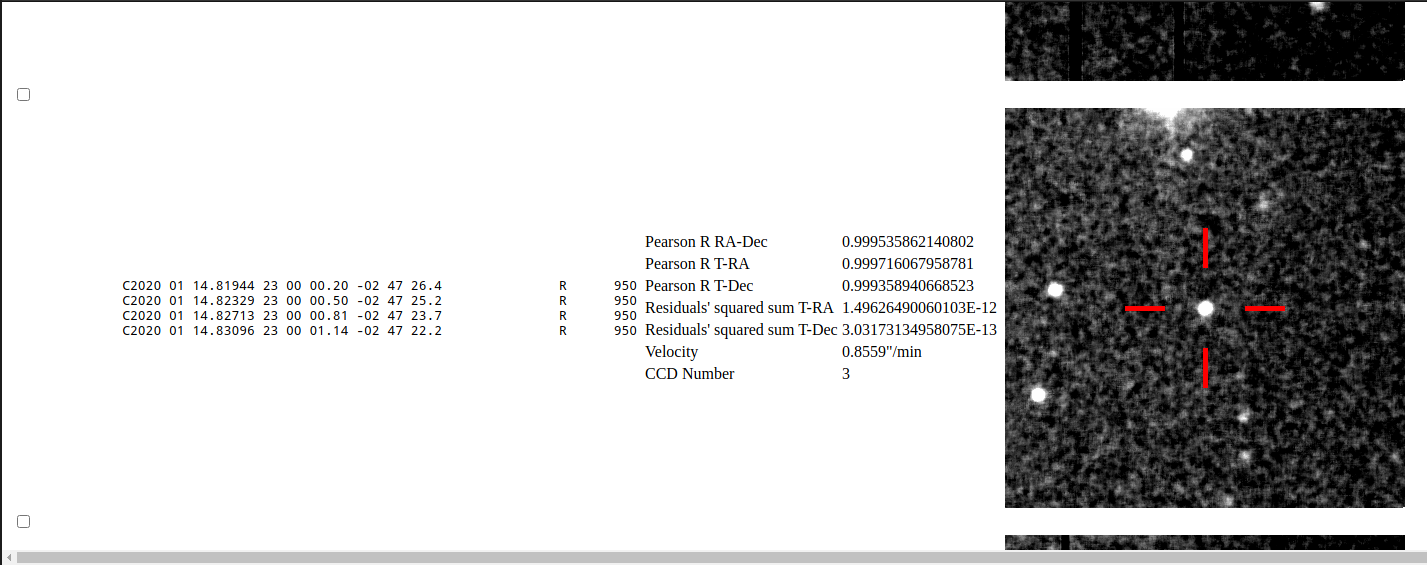}
	\caption{A valid detection as a reducer would see it}
	\label{fig:wbr-detect}
	\end{subfigure}
\begin{subfigure}{0.45\textwidth}
	\centering
	\includegraphics[width=.9\linewidth]{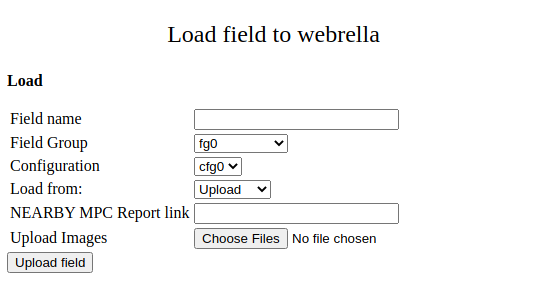}
	\caption{Uploading a field. It is possible to either directly upload the files or pull them from an existing NEARBY deployment residing on the same server.}
	\label{fig:wbr-load}
\end{subfigure}
\hspace{0.1\textwidth}
\begin{subfigure}{0.45\textwidth}
	\centering
	\includegraphics[width=.9\linewidth]{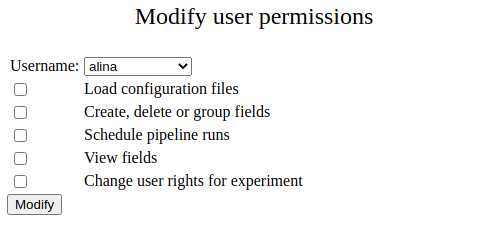}
	\caption{Adjusting the permissions of a Webrella user for a survey}
	\label{fig:wbr-perm}
\end{subfigure}
	\caption{Webrella interface}
	\label{fig:webrella}
\end{figure*}
\section{Umbrella2 core platform provisions}
Umbrella2 has a set of algorithms and interfaces on which the rest of the platform is based. They are enumerated below:
\subsection{Shared components}
A common library provides the types for Cartesian coordinates on 2D images, Projection plane coordinates (as in the FITS 3.0 standard \citep{fitsStandard}), and Equatorial coordinates, as well as the corresponding velocity vectors. Also present in this common library are the interface definitions for transforming between the different coordinate types. A converter to and from Minor Planet Center optical report format \cite{mpcobsformat} of the coordinates is also provided.\\

The common library also provides a plugin system (lightweight dependency-inversion), which can scan loaded modules and automatically use them in extensible components (such as the WCS projection types). Another extensibility feature present in the common library is the property model, a type-safe mechanism for dynamically attaching new data (which is usually either optional or user-defined) to objects implementing the property model (such as detections and tracklets).
\subsection{Image I/O}
Umbrella2 provides support for generic image types -- besides the I/O implementation for FITS files included in the I/O library, users may write support for other image formats of their own convenience. The FITS implementation in Umbrella2 handles the images according to the FITS Standard Version 3.0 \citep{fitsStandard}, including multi-image variants (MEF) and all 6 types of data representation (BITPIX), with extensible support for parsing headers and WCS projection types. There are however a few known missing features:\\
\begin{itemize}
	\item Currently the only the gnomonic (TAN) projection type has an implementation. (however, other projection types can be implemented in user plugins).
	\item No support for BSCALE and BZERO flux scaling.
	\item No support for quoting via consecutive single quotes in the headers.
\end{itemize}
The last two in particular make Umbrella2 non-compliant to the FITS standard. These shortcomings may be addressed in future releases.\\

A noteworthy detail is that Umbrella2 accepts FITS files either via memory-mapped files or from non-seekable streams, where the contents are being copied into memory, so that files may be streamed over networks if deemed necessary. For performance considerations, all images are by design read in chunks, with thread-safe accesses synchronized by a readers-writers lock allowing independent access to non-overlapping portions of the image. The lock can be disabled when access is proxied and the proxy provides the locking facility.
\subsection{Algorithms Framework}
To maintain a clean, high-level implementation of all image processing tasks while retaining high performance levels, Umbrella provides a framework that abstracts away image I/O and algorithm scheduling. This framework is, of course, exposed to library users, so that users may run their own algorithms on par with built-in ones. Included in the framework are also common definitions and interfaces that provide a high degree of modularity and seamless integration of different algorithms in pipelines.
\subsubsection{Detections and Tracklets}
The algorithms framework offers the types representing detections on individual images and tracklets (sets of detections that correspond to the same object on consecutive images). Both of these types are extensible through the property model. Very common properties are provided in the library: object identity, photometry measurements (flux and magnitude), the coordinates on the image (including all coordinates of the detection blob in the X-Y plane), shape and size information, pairing information, and velocity regression data.
\subsubsection{Object identity}
Umbrella2 can match MPC names of asteroids to tracklets, or create its own provisional names (given a field name, tracklet number, and optionally a CCD number). It supports both numbered and temporary designations for objects, as well as the packed form of the two (provisional names created by Umbrella are given in the form of packed temporary designations, where the first 4 characters are the field name). Name matching is implemented with a scoring algorithm, which gives scores normalized as integers from 0 to 100 (minimum score 1). The score takes into account the distance between the estimated and measured positions of the object and decreases exponentially with the number of detections that are very far from the estimated positions:
\[ \textbf{Score} = \frac{2^{\sqrt{n}}}{\bar{d} + 1} \]
normalized to $100$, where $n$ is the number of detections and $\bar{d}$ is the average distance between the detection and the SkyBoT prediction in arcseconds.
\begin{figure*}
	\centering
	\includegraphics[width=.9\linewidth]{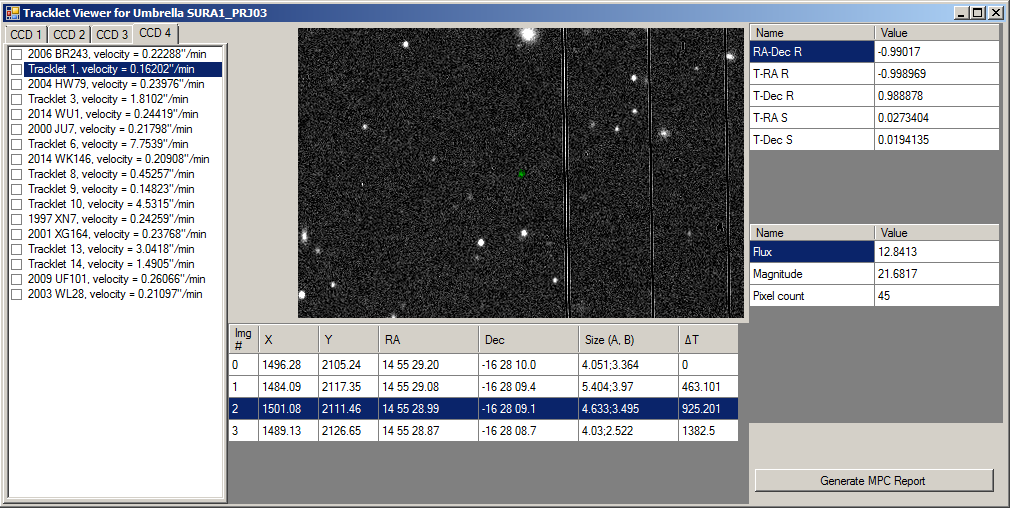}
	\caption[Tracklet Viewer]{An object that is not in the SkyBoT database.}
	\label{fig:vn-trackview-unk}
\end{figure*}
\subsubsection{Scheduler for high-level image algorithms}
Due to the amounts of data to be processed, a goal of Umbrella is for algorithms to make use of all available computing resources. This is achieved by a scheduler for the algorithms that runs the computation kernel in parallel over multiple threads, defaulting to one per logical core, on the CPU. Future versions of Umbrella may allow support for scheduling the algorithms on GPGPUs (General Purpose Graphics Processing Units) too. A notable feature of the scheduler is that it also separates the I/O layer from the algorithms, which allows algorithm developers to focus on the algorithms themselves.\\

It should be noted that the current CPU scheduler implementation comes with some limitations regarding input images where WCS crops are significant (such as when input images have lower overlap) as well as image chunk size and the level of parallelism.
\subsubsection{Miscellaneous}
Several convenience math routines are also provided with the algorithms framework for: linear regression, intersection of lines, and intersection of semilines with rectangles. Besides the math routines, there are also three data structures available for algorithms: a quad tree, a multi-threaded resource pool, and a graph structure for finding connected components.
\subsection{Visualization components}
\subsubsection{FITS Viewer component}
To present sections of FITS images within Umbrella, the software package includes a FITS viewer widget that can display images centered around a given coordinate. The viewer also supports highlighting a given set of pixels. Data is displayed via a scaling algorithm; currently, only a linear scale is implemented.
\subsubsection{Tracklet viewer}
\begin{figure*}
	\centering
	\includegraphics[width=.9\linewidth]{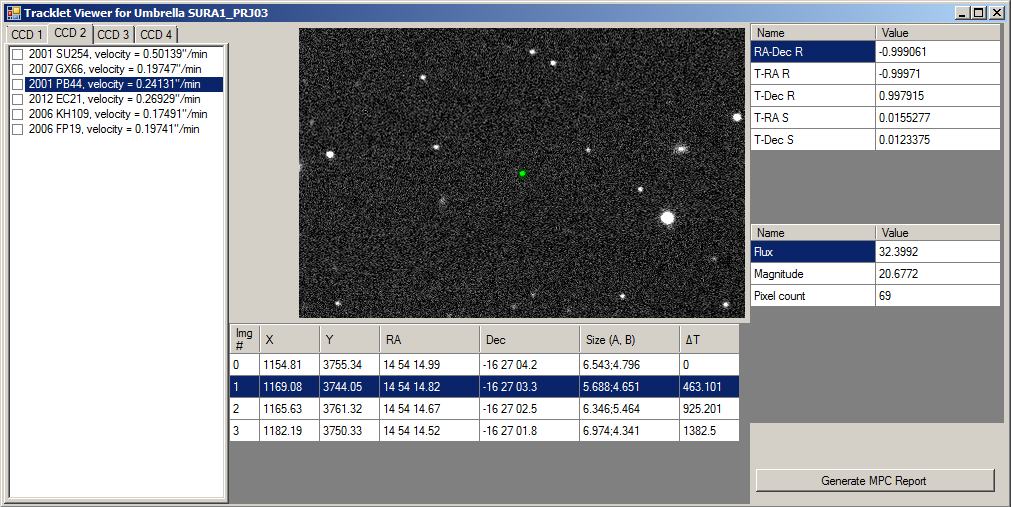}
	\caption[Tracklet Viewer]{Example of a high quality input image. Detections on each image are shown individually. Important information about the detection is available at a glance.}
	\label{fig:vn-trackview-hq}
\end{figure*}

\begin{figure*}
	\centering
	\includegraphics[width=.8\linewidth]{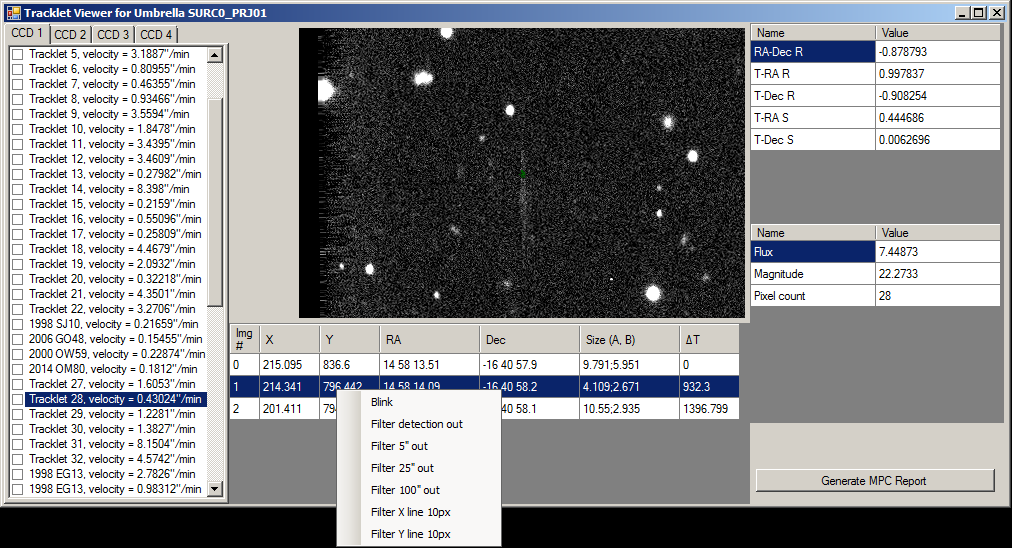}
	\caption[Tracklet Viewer]{On inputs where image artifacts greatly increase false detection rate, manual filtering is possible.}
	\label{fig:vn-trackview-filter}
\end{figure*}

The results of a pipeline using Umbrella can be displayed using a "Tracklet viewer" window (Figure \ref{fig:vn-trackview-hq}). This window offers visual and parametric inspection of the tracklets (and their individual detections). Visual inspection may be performed on any of the input or intermediate images (which have been tagged by the pipeline as corresponding to the same original image). The displayed parameters are selected from those computed in the pipeline. Wherever possible, names of the tracklets are obtained from their ObjectIdentity property(see Figure \ref{fig:vn-trackview-unk}). To better cope with the cases where the pipeline filtering might break down (due to missing or incorrect badpixel files, very bright stars, or broken optics correction), portions of the WCS or image coordinates can be manually filtered out from the results (see Figure \ref{fig:vn-trackview-filter}). Also for the convenience of the reducers, common operations (navigation, blinking) can be directly performed using the keyboard.
\section{Algorithms provided in the core repository}
Umbrella provides algorithms for all steps in the detection and processing of the moving objects:
\subsection{Noise removal}
Noise removal is performed using multiple algorithms for different purposes. Input images can be masked with a badpixel file. Initial per-image denoising can be performed with a trimmed mean filter (of which there exists a variant combined with the badpixel filter, which yields significantly better results). Cross-image denoising for creating a static objects mask can be done with a median filter. A further deep smoothing can be performed to use a much faster version of the AFAV long trail detection algorithm. There is also a simple implementation of image normalization.
\subsection{Object detection}
The library comes with two built-in object detection algorithms. The first is a simple threshold method, using two thresholds in a hysteresis flood fill. Historically, this is the algorithm used in the first version of Umbrella; it is referred to as the blob detection algorithm. The other detection algorithm uses a State Variable Auxiliary Feature Augmented Voting Hough Transform (SV-AFAV-HT, a technique based on the Hough Transform, paper by Stănescu in prep; a working description also presented in Annex A) on connectivity, which combines local and global properties using a Markov variable along the Radon line basis, to identify potential long trails. These potential long trails are further detected through another flood fill algorithm, the blobs being combined to yield more accurate trail detections. It is also possible to import Source Extractor \citep{sExtractor} catalog files.\\

There is also a variant of the blob algorithm for detection recovery on original images, which estimates positions from the tracklet velocity regression and can work with lower thresholds.
\subsection{Object pairing \& filtering}
Per-image detections are combined into tracklets through a pairing algorithm. There are two implementations currently available in Umbrella, one newer and which is expected to give better results (but has not had extensive testing on long trails) and an older one, better tested, but with a weaker design (which is why it is considered obsolete).\\

The library has a few built-in detection and tracklet filters. Currently, they are boolean, but this is expected to change in the future, so that the parameters can be automatically tuned through machine learning.
\subsection{External APIs and tools}
Umbrella can take advantage of existing platforms in reducing images. Currently, it hosts methods for accessing SkyBoT \citep{skyBot} and VizieR \citep{vizier} APIs and algorithms to integrate the query results into the pipeline: there are functions to match objects to SkyBoT name lists, stars to VizieR star lists and for calibrating the zero-point magnitude. Furthermore, there are functions for running the Digest2 \citep{digest2} software locally on the resulting MPC reports, so that objects of interest can be spotted quickly.
\section{Detailed description of the algorithms}
\begin{figure*}
	\begin{subfigure}{.5\textwidth}
		\centering
		\includegraphics[width=0.9\linewidth]{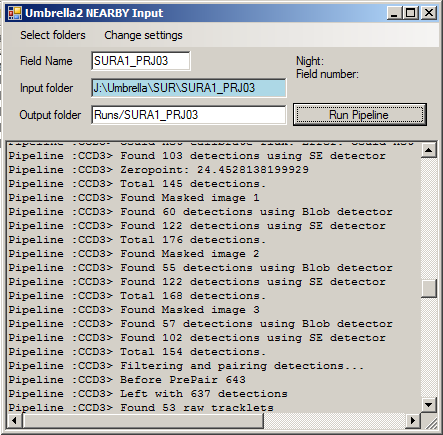}
		\caption{Example from the SURA1 field. Input folder is always validated before calling the pipeline. Example of the software running on Windows.}
		\label{fig:vn-suramain}
	\end{subfigure}%
	\begin{subfigure}{.5\textwidth}
		\centering
		\includegraphics[width=0.9\linewidth]{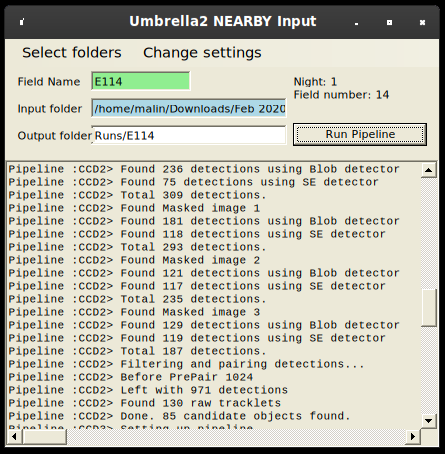}
		\caption{Fields named with the conventional EURONEAR name (Enff) are recognized as such by ViaNearby and can make use of additional defaults (such as default names compatible with the MPC Optical Report format for new detections). Example of the software running on Linux.}
		\label{fig:vn-namedfield}
	\end{subfigure}
	\begin{subfigure}{.5\textwidth}
		\centering
		\includegraphics[width=0.9\linewidth]{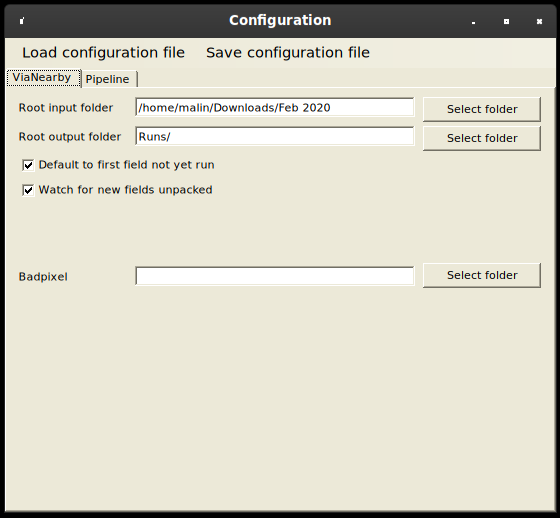}
		\caption{ViaNearby options. If properly set up, all form fields are filled automatically with the first field not yet run, and the user only has to tab-cycle them for validation before running the pipeline.}
		\label{fig:vn-conf2}
	\end{subfigure}
	\begin{subfigure}{.5\textwidth}
		\centering
		\includegraphics[width=0.9\linewidth]{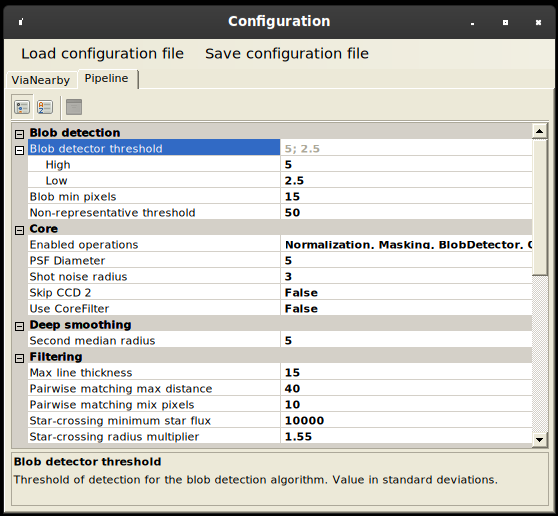}
		\caption{Configuration of the pipeline parameters.}
		\label{fig:vn-conf}
	\end{subfigure}
	\caption{ViaNearby desktop pipeline}
	\label{fig:vn-shots}
\end{figure*}
\subsection{Noise removal}
All the trimmed mean and median algorithms, except for the single-image median, use the built-in sorting functions of the environment \citep{arraySort} (insertion sort for small array and quicksort for larger ones). The single-image median uses a custom quickselect, where initial pivots are estimated from the mean of the input pixels and the last output pixels.\\

Initial pivot choice algorithm:
\begin{itemize}
	\item A linear median estimate is taken from the last 2 median values
	\item The previous estimate is averaged with the mean of the kernel window
	\item Upper pivot is estimate + $0.3$ standard deviations
	\item Lower pivot is estimate - $0.35$ standard deviations
	\item If median is not in range, take upper or lower pivot at estimate $\pm 1$ standard deviation 
\end{itemize}
The parameters were obtained via a quick empirical benchmark on typical Umbrella input images.
\subsection{Normalization}
The current normalization algorithm is a quick workaround for removing the halos of very bright stars; it is expected that this algorithm will be deprecated and replaced by a better one in the future, as it has obvious flaws.\\

The algorithm segments the image into square regions and computes the local median. This is later subtracted from the image using the following formulas, where $\Omega = \left\{ (-,-), (-,+), (+,-), (+,+) \right\}$ is the set of the 4 surrounding regions:
\[ d_\textbf{Sum} = \sum_{\iota \in \Omega} d_\iota \]
\[ M_\textbf{Interpolated} = \sum_{\iota \in \Omega} (d_\textbf{Sum} - d_\iota) \times M_\iota \]
\[ O = I - M_\textbf{Interpolated} \]
$M_\iota$ is the local median of the region, and $d_\iota$ is the distance in pixels from the current pixel to the region.
\subsection{Detection algorithms}
\subsubsection{Blob}
The blob detection algorithm performs a standard flood fill, with a 4-way connected component recognition, and using a 2-threshold hysteresis for robustness.
\subsubsection{Long Trails}
The long trail detection algorithm is based on the Auxiliary Feature Augmented Voting Hough Transform (a detailed description is to be published in a separate paper). The auxiliary feature used for this algorithm is the local connected-ness of the trails, which is computed efficiently using a state variable along the Radon voting basis. The vote table is then thresholded and high scores are searched with another blob-like algorithm, which can also pair neighboring disjoint detections.\\

The default settings for scheduling the algorithm are to split the input image into $300 \text{px} \times 300 \text{px}$ tiles that overlap with $50 \text{px}$ margins (on each side). These settings may be used as-is for scheduling or may be adapted by the pipeline. It should be noted that the reference pipeline uses the default values and does not expose these settings to the end user.
\subsection{Detection pairing}
The newer pairing algorithm provided in Umbrella works by linearly estimating the positions of detections and pairing accordingly. This is achieved by putting all detections in a quad tree, and for each pair of detections, performing the following operations:
\begin{itemize}
	\item Performs a quick check for pair compatibility\\
	Here it is checked that the detections are from different images and that the distance between them is not too large (given their size on input images, the exposure time and the time difference between images).
	\item Estimates positions of the supposed object on all images, along with an estimate of the position error
	\item Queries the quad-tree for all detections around the estimated position that are within the position error at the correct time
	\item For each detection that matches the query, performs a linear fit of the coordinates of the initial pair and the match against time (this is done via ordinary least squares, as time is assumed to be accurately measured and thus exogenous to the model)
	\item If detections are matched on at least one other image (thus 3 images in total), creates a tracklet.
\end{itemize}
\subsection{Filtering}
Filtering is done through individual pass/drop filters, applied both to image detections and tracklets. Several filters are included in the Umbrella2 library. Another collection of filters is implemented in the reference pipeline.
\section{The reference pipeline}
The current deployments of Umbrella-based software include a standalone desktop application (screenshots in Figure \ref{fig:vn-shots}) and a web-based server solution (screenshots in Figure \ref{fig:webrella}). Both implementations currently use a common pipeline, but this is expected to change in the future. The following describes the 7 steps performed in this reference pipeline (also shown in Figure \ref{fig:pipeline-flowchart}):
\begin{figure*}
	\centering
	\includegraphics[width=.8\linewidth]{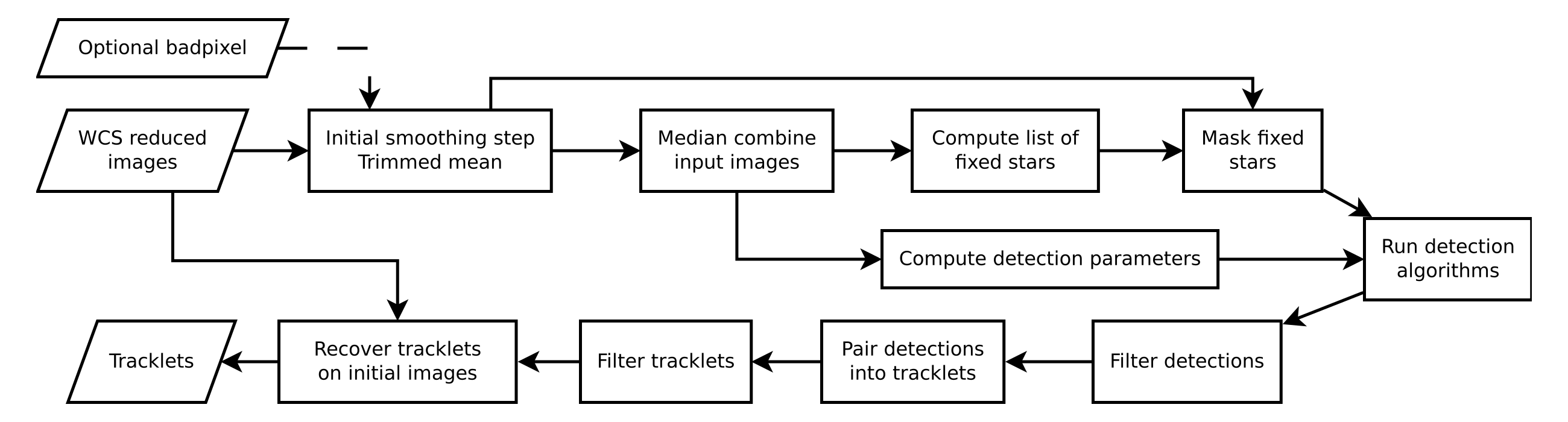}
	\caption{Flowchart of the Umbrella reference pipeline}
	\label{fig:pipeline-flowchart}
\end{figure*}
\subsection{Initialization}
On calling the pipeline, the first steps are:
\begin{itemize}
	\item Enabling or disabling scaling the image brightness according to SWarp headers.
	\item Checking whether the temporary directory exists (and ensuring it does)
	\item Generating the kernels used in the image processing functions ahead
	\item Checking whether a badpixel file is provided and reading the badpixel data
\end{itemize}
\subsection{Image processing}
There are three steps of image processing in the reference pipeline. These are:
\subsubsection{Initial smoothing step}
In the initial smoothing step, the images are first noise filtered with a trimmed mean filter (either the usual one or the badpixel-aware one, depending on whether the badpixel file is provided). Optionally, a normalization step is performed directly after.
\subsubsection{Median combine}
A median image is generated by stacking the previously processed images. The median image is used to compute the required parameters for the detection algorithms (such as thresholds). A list of fixed stars is also generated at this step.
\subsubsection{Removal of fixed stars}
The images processed in the pre-median step are masked with the list of fixed stars in the median. Following the masking, an optional deep smoothing step is performed, which is designed to provide smooth data for connected-ness recognition in the long trail algorithm.
\subsection{Detection and filtering}
The reference pipeline currently uses three detection algorithms, two internal (Blob and Long trail) and one external (Source Extractor, with catalogs imported by Umbrella). The resulting detections are filtered, paired, and filtered again. Finally, the tracklets are recovered on the input images (using the estimated positions from a linear fit) and displayed to the reducers.
\section{Results}
\begin{figure*}
	\centering
	\includegraphics[width=0.9\textwidth]{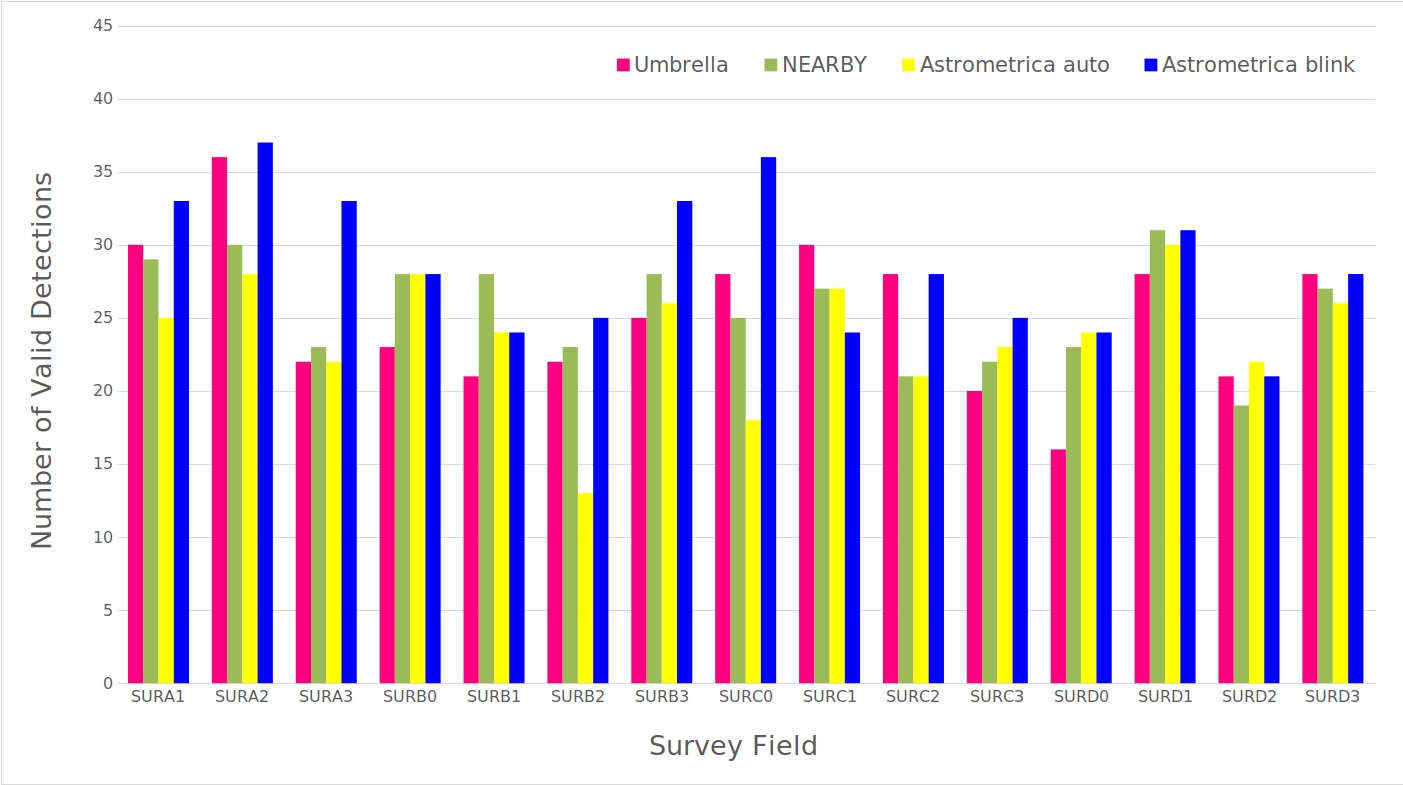}
	\caption{Umbrella compared to other automated detection software}
	\label{fig:defSettings}
\end{figure*}
\subsection{Test setup}
A high quality dataset created for the similar NEARBY project (also within EURONEAR) was used to assess the detection rate of Umbrella. The exposures were taken with the Wide Field Camera on the Isaac Newton Telescope, close to opposition. The average seeing of the exposures is approx. 1.5". The results are from the latest release of ViaNearby at the time of comparison.
The results from Figure \ref{fig:defSettings} use the default settings, which are presented in Table \ref{tab:defSettings}. These settings have been selected over time to offer an even balance between false negative and false positive results on most images captured by the EURONEAR surveys.
\subsection{Summary of results}
Umbrella has detected 378 detections in the 15 WFC fields (4 sq. deg.), compared to 357 detected by Astrometrica (106\%), 384 detected by NEARBY (98\%) and 430 by manual blink (88 \%). Despite the lower detection rate, it has detected 13 asteroids not detected by NEARBY, one of which missed also by manual blink.\\

There have been 1826 false positives, for a total true positive rate of 17\%. However, this rate has shown tremendous variation between different fields and CCDs; for example, the SURA1 CCD2 has had 6 true positives and no false positives, while SURD0 CCD3 has had 2 true positives and 106 false positives.
\subsection{Detection hijacking}
It has been noticed that on the test data (from the logs of the object identification component) that on some fields up to an additional 10 \% (compared to the detection rate shown in this paper) known objects (from the SkyBoT database) should have been detected if not for detection hijacking, described below. It is unknown how many detections that do not show up in the SkyBoT database have been missed in this way.\\

One of the design objectives of the current pairing algorithm was that tracklets shall not be duplicated. Therefore, the earliest steps in the pairing algorithm check if the chosen pair have been already included into a tracklet. When a large number of false detections are present, it becomes more likely that true positive are paired with false positives. Let us consider a field on which an object was detected on $n$ exposures. If more than $n-2$ of these detections are paired with false detections before the true positives are paired, then the true detections will no longer be paired, and the object is lost by Umbrella.\\

For the 4-exposure fields of the NEASUR dataset described here, if 2 detections are paired with a false detection, or more commonly, when the object is detected only on 3 images (which can happen due to poor badpixel removal or involvment with the PSF of a star), if 1 (of any 3) detection is paired incorrectly, the object is lost. This effect is particularly relevant when images have extended bad areas (which is the case for one of the CCDs in the NEASUR dataset) and less than ideal badpixel removal methods, where bad areas not only can obscure a detection, but also introduce artifacts that can hijack the other detections.\\

Finaly, this phenomenon is also relevant when operating at high sensitivity (low SNR threshold for detection), such as when one desires a more "brute force" approach to asteroid detection. In such cases, lowering the thresholds for detection quickly increases false positives, which then results in greatly increased chances of hijacking. This in fact creates a barrier to the number of achievable object discoveries for a given image.
\subsection{Comment on results}
The settings used to reduce the images are the software default, also used during EURONEAR real-time surveys. While not optimized for these particular images and not selected for the maximum detection rate, these settings do yield most of the detections currently achievable with Umbrella (likely above 80 \% maximum detection rate). In particular, due to detection hijacking, greatly increasing false positives can have a detrimental effect on detecting true positives.\\

It should be noted that these results represent a snapshot of the performance of the current implementation of the pipelines, with adjustments to algorithms, pipeline and settings being common. To automate this process, the authors have envisioned a software tool that can track detections (known or unknown to Umbrella) across the pipeline on a large scale and provide statistics and visualizations such that weaknesses in algorithms can be identified and corrected, while detection settings can be tweaked more appropriately. Partial support for such a tool has been added to the reference pipeline implementation (OutputRemovedDetections in Core/Enabled operations - see Table \ref{tab:defSettings}) and it is hoped that the tool is ready in the near future.\\
\section{Acknowledgments}
The authors would like to thank Daniel Berteșteanu and Costin Boldea for testing Webrella and suggesting improvements to the interface, Alexandru Georoceanu for testing the desktop version (especially for reporting that the desktop interface is broken on 64-bit MacOS), and Elisabeta Petrescu for cross-validating difficult detections in early live tests.
\begin{table*}
	\centering
	\texttt{
		\begin{tabular}{|l|c|}
			\hline
			\textbf{Setting} & \textbf{Value}\\
			\hline
			Star masking / Star masking threshold & 3.5; 2\\
			Star masking / Extra mask radius & 1\\
			Star masking / Mask radius multiplier & 1.1\\
			Blob detection / Blob detector threshold & 5; 2.5\\
			Blob detection / Blob min pixels & 15\\
			Blob detection / Non-representative threshold & 50\\
			General pipeline properties / Standard BITPIX & -32\\
			General pipeline properties / Maximum algorithm detections & 1000\\
			Core / Shot noise radius & 3\\
			Core / Use CoreFilter & false\\
			Core / Skip CCD 2 & false\\
			Deep smoothing / Second median radius & 5\\
			Input / Correct SWARP & true\\
			Normalization / Normalization Mesh size & 40\\
			Core / PSF Diameter & 5\\
			Long trails / RLHT Threshold & 10\\
			Long trails / Segment selection threshold & 5; 2.5\\
			Long trails / Maximum interblob distance & 40\\
			Long trails / Minimum trail pixels & 100\\
			Filtering / Max line thickness & 15\\
			Filtering / Pairwise matching max distance & 40\\
			Filtering / Pairwise matching mix pixels & 10\\
			Filtering / Star-crossing radius multiplier & 1.55\\
			Filtering / Star-crossing minimum star flux & 10000\\
			Pairing / Same-object separation & 0.1\\
			Pairing / Max residual sum & 2\\
			Pairing / Extra search radius & 3.5\\
			Core / Enabled operations & Norm, Mask, BD, ORD, SE\footnotemark\\
			Original image recovery / Detection threshold & 1.75\\
			Original image recovery / Recovery radius & 10\\
			Reporting / Observatory code & 950\\
			Reporting / Magnitude Band & R\\
			Reporting / SkyBoT Radius & 5\\
			\hline
		\end{tabular}
	}
	\caption{Default Umbrella settings}
	\label{tab:defSettings}
\end{table*}
\footnotetext{Normalization, Masking, BlobDetector, OutputRemovedDetections, SourceExtractor}

\pagebreak
\appendix
\section{Working description of the State Variable Auxiliary Feature Augmented Voting Hough Transform}
The SV-AFAV-HT is a modification of the Hough Transform to improve detection rate through the use of an auxiliary feature of the target objects. This is particularly important for detection of faint trails, where unwanted spurious sources (such as faint stars under the masking threshold) can yield stronger responses than asteroid trails.\\

The SV-AFAV-HT follows the outline of the classic (rho-theta, as in \citet{hough-dudahart}) Hough Transform, i.e. scans the image along the Radon basis and keeps a voting table which is updated with a vote from each pixel; however, unlike the classic Hough Transform, SV-AFAV-HT gives different weights to the votes of each pixel. The auxiliary feature chosen for line streaks such as asteroid trails is connectivity (as asteroid trails should be continuous up to image noise). For this purpose, a sliding window mean (of configurable length) is used to derive the weighing factors in $O(1)$ time complexity. This is achieved by using per-line state that evolves as a Markov chain - the steps through which the per-pixel weight is obtained are described below:
\begin{itemize}
	\item The windowed mean is updated (a ring buffer is used to store past values along the Radon basis, so that the update is $O(1)$)
	\item Scale the windowed mean according to the user-configured threshold, resulting in $M_S$.
	\item Apply a positive clamp to the scaled mean to obtain $M_C$.
	\item Obtain the decay factor $\delta$ using the homographic function $\frac{x}{1+x}$ on $M_C$, scaling the result between the two user-defined limits
	\item Compute a raw weighing value by decaying its previous value with $\delta$ and adding $M_S$
	\item Obtain the pixel weight by scaling the raw weight with an atan function into the interval $0.5-6.5$.
\end{itemize}
The initial state is obtained by computing the mean from the first window and filling in the raw weighing value with the mean.
\pagebreak

\bibliographystyle{model2-names}
\bibliography{bibliography.bib}

\end{document}